\documentclass[conference]{IEEEtran}
\IEEEoverridecommandlockouts
\usepackage[final]{changes}  
\setaddedmarkup{\uline{#1}}  
\usepackage{booktabs} 
\usepackage{cite}
\usepackage{amsmath,amssymb,amsfonts}
\usepackage{algorithmic}
\usepackage{graphicx}
\usepackage{textcomp}
\usepackage{xcolor}
\usepackage{algorithm}
\usepackage{array}
\usepackage[caption=false,font=normalsize,labelfont=sf,textfont=sf]{subfig}
\usepackage{stfloats}
\usepackage{url}
\usepackage{verbatim}
\usepackage{xcolor}
\usepackage{textgreek}
\usepackage{colortbl}
\usepackage{subcaption}
\usepackage{graphicx}
\usepackage{fancyhdr}
\usepackage{multirow}
\usepackage{booktabs}
\usepackage{array}
\usepackage{tabularray}
\usepackage{threeparttable}
\graphicspath{{figure/}}

\def\BibTeX{{\rm B\kern-.05em{\sc i\kern-.025em b}\kern-.08em
    T\kern-.1667em\lower.7ex\hbox{E}\kern-.125emX}}
\usepackage{fancyhdr}
\usepackage{hyperref} 
\hypersetup{hidelinks}
\fancypagestyle{ieee}{ 
    \fancyhf{} 
    \fancyhead[C]{
    \footnotesize This work has been accepted for publication in \textbf{IEEE Journal on Exploratory Solid-State Computational Devices and Circuits}. This is the author’s version, which has not been fully edited, and content may change prior to final publication. Citation information: \textbf{DOI 10.1109/JXCDC.2026.3680281}}
    \fancyfoot[C]{
    \footnotesize \textbf{\copyright~2026 IEEE}.  Personal use of this material is permitted.  Permission from IEEE must be obtained for all other uses, in any current or future media, including reprinting/republishing this material for advertising or promotional purposes, creating new collective works, for resale or redistribution to servers or lists, or reuse of any copyrighted component of this work in other works.}

}
\begin{document}
\title{OISMA: On-the-fly In-memory Stochastic Multiplication Architecture for Approximate Matrix-Multiplication}

\author{
Shady Agwa \textit{\IEEEauthorrefmark{1}}, Yihan Pan \textit{\IEEEauthorrefmark{1}}, 
Georgios Papandroulidakis and Themis Prodromakis
\\
Centre for Electronics Frontiers, Institute for Integrated Micro and Nano Systems, \\
School of Engineering, The University of Edinburgh, UK.\\
Corresponding Author: Shady Agwa (e-mail: shady.agwa@ed.ac.uk).
\thanks{
\textit{\IEEEauthorrefmark{1} Shady Agwa and Yihan Pan contributed equally to this work.}
}
\thanks{This work was supported by EPSRC FORTE Programme (Grant No. EP/R024642/2) and by the RAEng CiET (Grant No. CiET1819/2/93).}
}

\maketitle

\thispagestyle{ieee} 
\begin{abstract}
Artificial Intelligence models are currently driven by a significant up-scaling of their complexity, with massive matrix-multiplication workloads representing the major computational bottleneck. In-memory computing architectures are proposed to avoid the Von Neumann bottleneck. However, both digital/binary-based and analogue in-memory computing architectures suffer from various limitations, which significantly degrade the performance and energy efficiency gains. This work proposes OISMA, an energy-efficient in-memory computing architecture that utilizes the computational simplicity of a quasi-stochastic computing domain (Bent-Pyramid system), while keeping the same efficiency, scalability, and productivity of digital memories. OISMA converts normal memory read operations into in-situ stochastic multiplication operations with a negligible cost. An accumulation periphery then accumulates the output multiplication bitstreams, achieving the matrix multiplication functionality.
A 4KB 1T1R OISMA array was implemented using a commercial 180nm technology node and in-house RRAM technology. At 50 MHz, it achieves 0.789 TOPS/W and 3.98 GOPS/mm$^2$ for energy and area efficiency, respectively, occupying an effective computing area of 0.804241 mm$^2$. Scaling OISMA to 22nm technology shows a significant improvement of two orders of magnitude in energy efficiency and one order of magnitude in area efficiency, compared to dense matrix multiplication in-memory computing architectures.
\end{abstract}

\begin{IEEEkeywords}
AI, In-Memory Computing, Stochastic Computing, RRAM, Unconventional Computing.
\end{IEEEkeywords}
\section{INTRODUCTION}

\IEEEPARstart{A}{rtificial} Intelligence (AI) has revolutionized our daily lives in various domains such as security and defense, healthcare, automotive, robotics, and social media. However, the computational complexity of emerging AI models is continuously up-scaled, primarily driven by Large Language Models (LLMs) using Transformers, and Convolutional Neural Networks (CNNs). These massive-scale models incorporate hundreds of billions of parameters and perform computationally intensive operations as major workloads, mainly Matrix Multiplication ($MatMul$) and Convolution ($Conv$). Consequently, new critical challenges have emerged due to the ever-growing demands of performance and energy efficiency. Meanwhile, Moore’s Law is coming to an end due to the physical limits of device scaling, taking into consideration that LLMs are being scaled by orders of magnitude higher than Moore’s Law. On the other hand, Von Neumann architectures (i.e., CPUs and GPUs) were not originally designed for data-centric computing with massive datasets. This evolving scenario highlights three major bottlenecks that pose significant challenges for the future of AI~\cite{agwa2023digital,agwa2023bent}. Firstly, the Inter-Die bottleneck arises from the expenses of data transmission from one die (memory chip) to another (processing chip). Secondly, the Intra-Die bottleneck involves the still-costly data movement within the same die, such as manycore-like architectures with complex on-chip memory hierarchy. Thirdly, the Intra-Core bottleneck belongs to the computational cost of the Multiply-and-Accumulate (MAC) operation in the binary computing domain. 



In this paper, we introduce OISMA, an On-the-fly In-memory Stochastic Multiplication Architecture for deterministic approximate matrix multiplication. OISMA utilizes RRAMs to build digital 1T1R on-chip memory arrays that convert normal read operations into stochastic multiplications, achieving an energy efficiency of 0.789 TOPS/W at 180nm technology (79.22 TOPS/W if scaled to 22nm). An extensive $MatMul$ benchmarking using the quasi-stochastic Bent-Pyramid system~\cite{agwa2023bent} shows an average relative Frobenius error down to 1.81\%  compared to the 64-bit double precision floating point format.


\section{IN-MEMORY COMPUTING (IMC) DOMAINS}
Analog In-Memory Computing (IMC) architectures have been widely proposed to overcome the Von Neumann bottleneck using emerging technologies such as RRAMs~\cite{Yao_nature2020,Wan_nature2022,Kim_jetcas2022}. 
Analog IMC utilizes resistive crossbar arrays to perform $MatMul$ workloads in the analog domain through Kirchhoff's current law. Although multiplication and accumulation operations are easily performed in the analog domain, analog IMC architectures still inherit the same complexity and challenges of the analog design~\cite{Adam_nature2018,Liu_isscc2020}. The first challenge is the variability of analog memristors, including both device-to-device variations and cycle-to-cycle variations within the same device. Furthermore, memristor devices suffer from resistance drifting over time, which requires a continuous calibration process. This adds more constraints on the writing and reading circuitry, significantly increasing its overhead. The second challenge is the complexity of the analog/digital cross-domain interfacing, where Digital-to-Analog Converters (DACs) and Analog-to-Digital Converters (ADCs) are required with accurate-enough specifications for input/output conversion. Although analog IMC architectures run at very low supply voltage, to achieve high energy efficiency, the current-based computation and the complexity of the analog/digital cross-domain interfacing lead to long latencies and low throughput. Analog IMC also suffers from low computational/memory density due to the high complexity of the peripheral circuitry. Eventually, analog IMC approaches are still struggling with design scalability from system-level scale and technology node perspectives; Meanwhile, AI models are being scaled continuously with billions of parameters, requiring a faster and reliable hardware implementation approach that considers time-to-market, scalability, and compatibility with the software community.

In another direction, digital binary-based IMC architectures are also proposed to mitigate the data-movement bottleneck while inheriting the same productivity and scalability of digital architectures. 
These emerging architectures leverage the massive parallel resources of the memory structure to process data where it exists in DRAMs~\cite{seshadri-micro2013,Farahani-hpca2015}, RRAMs~\cite{agwa_iscas2022}, or SRAMs~\cite{Jeloka-vlsic2015,Eckert-ISCA2018,Fujiki-ISCA2019,Al-Hawaj-iscas2020,Al-Hawaj-hpca2023} using a bit-line computing approach. This in-situ computing approach activates two memory wordlines simultaneously, using two parallel address decoders, to achieve bit-wise logical operations (AND and NOR) between the two activated wordlines. Then, a near-memory logic stack is tightly coupled to the memory peripherals to perform more complex operations, such as additions and multiplications; however, at the cost of both throughput and energy efficiency. Digital binary-based IMCs consume hundreds of cycles to perform a MAC operation due to the complexity of their micro-algorithms~\cite{Al-Hawaj-iscas2020}. Consequently, these architectures failed to meet the expectations of high throughput and energy efficiency due to the computational complexity of the binary-based MAC operations.

Stochastic Computing (SC)~\cite{Alaghi_acm2013,Alaghi_dac2013,Alaghi_date2014,Winstead_springer2019,Groszewski_isqed2019,Alaghi_phd2015,Zhang_snw2019,Salehi_tvlsi2020,Alaghi_tcadics2018,Zhang_tcasii2020} is proposed as a middle ground between the digital (binary-based) and analog computing domains. Inspired by the spiking communication of biological neurons, stochastic data representations combine the analog’s computational simplicity (especially for multiplication and addition) with the high performance, robustness, productivity, and scalability of the digital domain. Therefore, SC unlocks the momentum of IMC by digitizing the neuromorphic computing while eliminating its cross-domain interfacing bottleneck~\cite{agwa2023digital,agwa2023bent}. SC numbers are represented by randomly generated bitstreams of ones and zeros, where these bitstreams are mapped to probabilities (from 0.0 to 1.0). Regarding unipolar stochastic numbers, the accompanied probability is calculated by the ratio of the ones to the total number of bits in the bitstream~\cite{Alaghi_phd2015}. For unipolar SC bitstreams, the SC multiplication is achieved by a bit-wise AND operation. 

\begin{figure}[!t]
    \centering
    \includegraphics[width=\columnwidth]{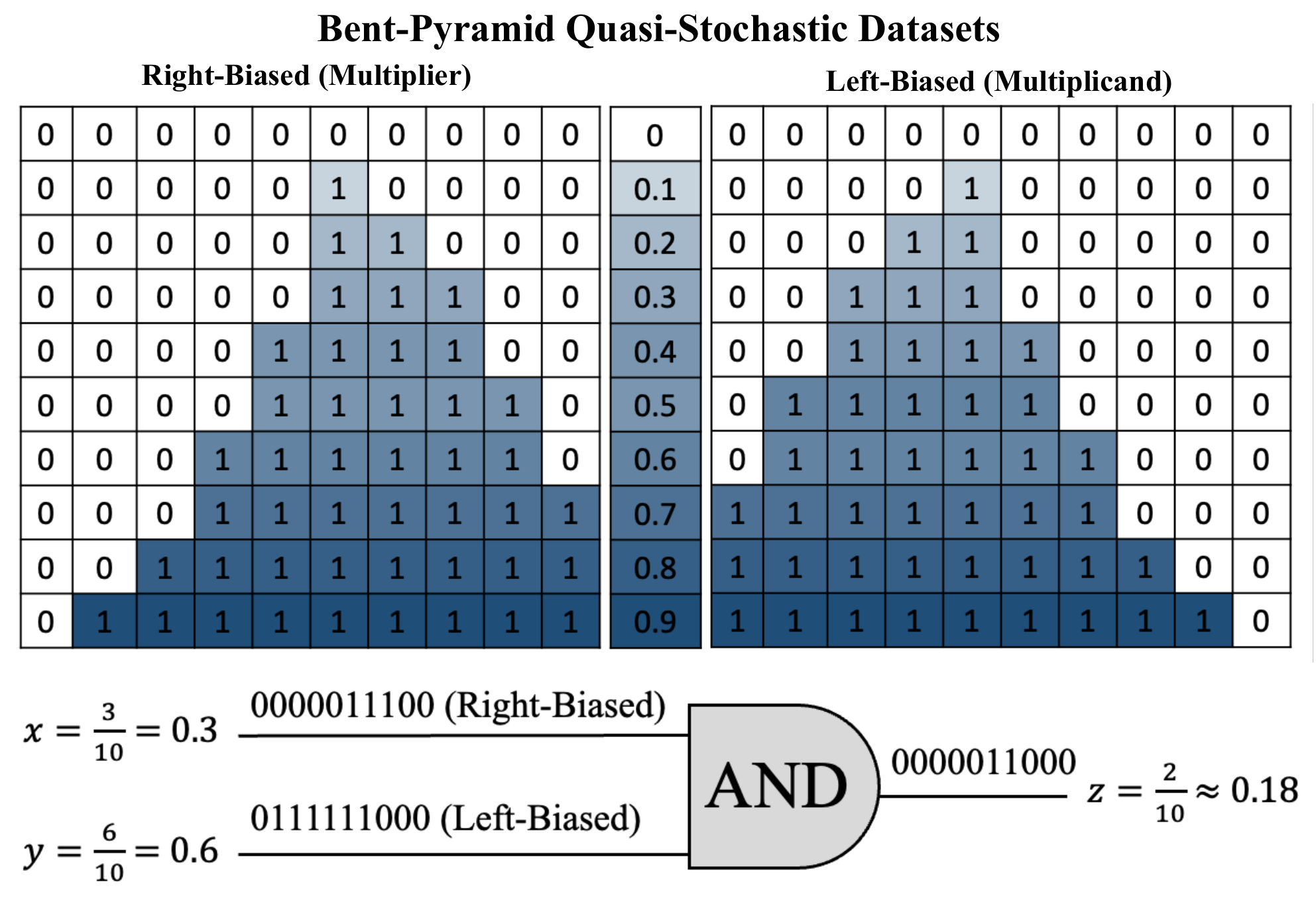}
    \caption{Bent-Pyramid bitstreams~\cite{agwa2023bent} (BP10), a stochastic multiplication example using the two different complementary datasets.
    }
    \label{fig:sc_and}
\end{figure}

Bent-Pyramid (BP) system is a new quasi-stochastic data representation that avoids the main pitfalls of conventional SC bitstreams~\cite{agwa2023bent}. The BP system has two complementary datasets of fixed 10-bit probability representations from 0.0 to 0.9 with 10\% step resolution. These two fixed complementary datasets of bitstreams are carefully designed to be uncorrelated, thereby aiming at maximizing the stochastic multiplication accuracy. Although the 10-bit BP system has deterministic input and output values, it still inherits the same computational simplicity of conventional SC multiplication. Unlike SC, BP system doesn’t require different seeds or random-number generation circuits. Moreover, BP numbers are generated within a single cycle, significantly improving latency and energy consumption per number generation in comparison to the conventional SC domain. Figure~\ref{fig:sc_and} shows the two complementary datasets of the 10-bit BP system, where probabilities from 0.0 to 0.9 are represented by right-biased and left-biased bitstreams. This different biasing minimizes the correlation between multipliers and multiplicands. 

\section{BENT-PYRAMID (BP) BENCHMARKING}

To investigate the accuracy of the 10-bit BP system, various $MatMul$ benchmarks were conducted, comparing 10-bit Bent-Pyramid (BP10) and 8-bit Floating Point (FP8) versus a baseline of 64-bit double-precision Floating Point (FP64). 
This accuracy analysis focuses on the format of E4M3 for FP8, with a 4-bit exponent and a 3-bit mantissa, which is widely proposed in the AI field~\cite{micikevicius2022fp8}.
Ignoring the negative sign for simplicity, FP8 offers 56 distinctive values between 0.0 to 1.0, achieved through its exponential nonlinearity. In contrast, BP10 provides only 10 distinctive values, utilizing a linear quantization structure.
\begin{figure*}[!t]
    \centering
    \includegraphics[width=\textwidth]{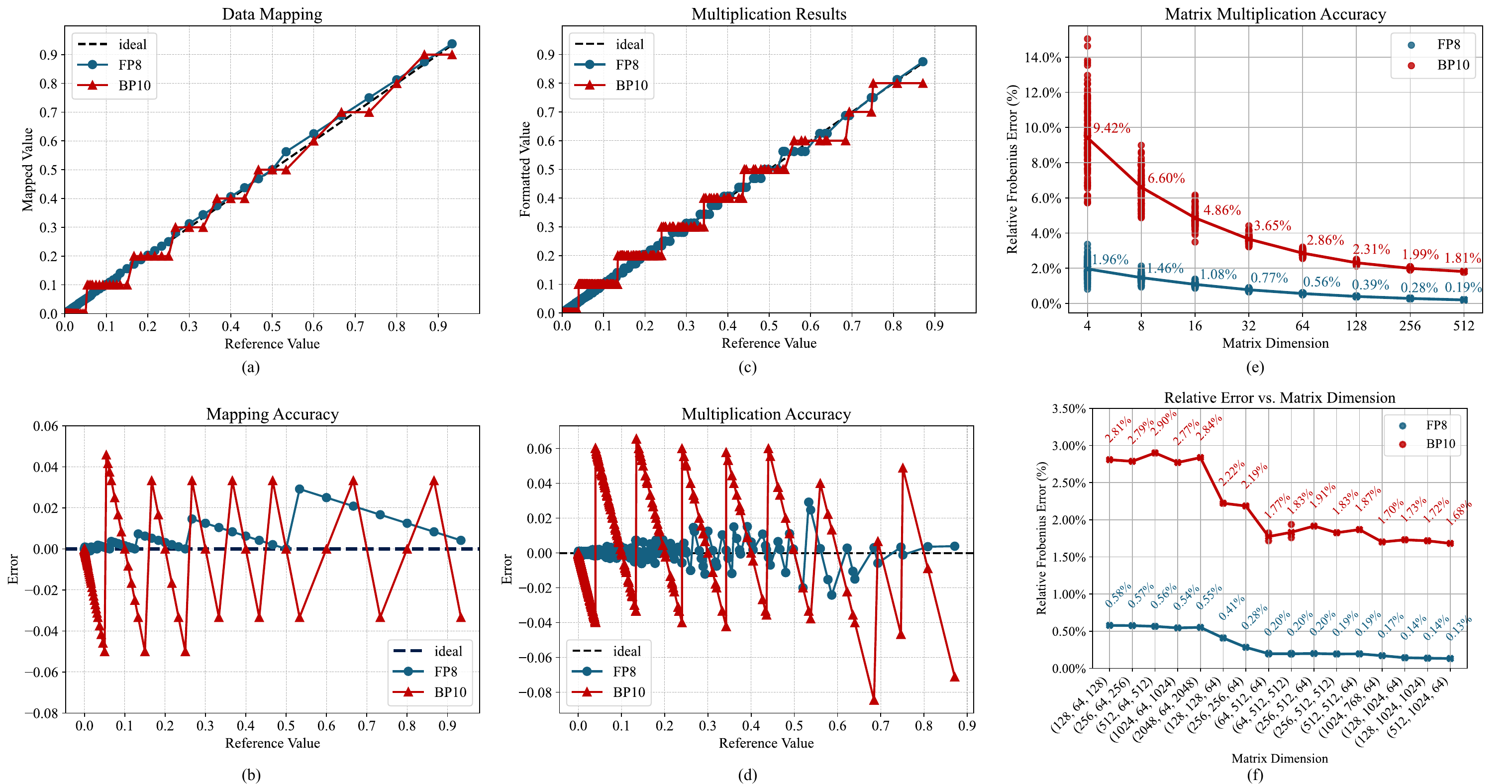}
    \caption{(a,b) Data mapping and mapping accuracy from normalized FP8 values (0.0 to 1.0) in \added{ideal }FP64 format to the valid equivalent FP8 and BP10 values. (c,d) Multiplication results and accuracy in FP8 and BP10 versus \added{ideal }FP64. 
    (e,f) Matrix-Multiplication benchmarking accuracy results for various matrix dimensions from 4x4 to 512x512 \added{with 100 independent random trials per dimension; }and for matrix dimensions that are commonly used in AI workload\deleted{, comparing FP8 and BP10 to }\added{. The Frobenius error of FP8 and BP10 matrix are computed relative to the ideal }FP64 format\deleted{.}\added{ labelled with average relative Frobenius error per test cases.} }
    \label{fig:benchmark}
\end{figure*}
Motivated by neural network processing, where data normalization is widely used for faster and more stable training, the positive range of FP8, from 0 to 240, is normalized to fall between 0.0 and 1.0, using FP64 format. The FP64 values, considered as the ideal baseline, are then mapped to their nearest FP8 and BP10 numbers. Figure~\ref{fig:benchmark}(a,b) shows the data mapping and the mapping accuracy of the two data formats. FP8 shows an average absolute error of 0.21\%, while BP10 reports an average absolute error of 1.19\%, which is 5.66x higher, considering that FP8 offers 5.6x more distinctive values than BP10. However, the mapping accuracy reveals that FP8 tends to produce only positive errors, while BP10 exhibits a fluctuating trend with positive and negative errors. These positive and negative mapping errors may lead to a lower accuracy-loss for some arithmetic operations.


For the multiplication process, the ideal data values are multiplied using the FP64 format and then normalized to the range of 0.0 to 1.0 using the max-min normalization. The FP8 multiplication results are obtained by multiplying FP8 values, and the outputs are then quantized from 64-bit to the nearest FP8 value.
Regarding the BP10 multiplication, right-biased and left-biased BP datasets in Figure~\ref{fig:sc_and} are used for multiplicands and multipliers to avoid input correlation. The generated BP10 numbers from the data mapping phase utilize bit-wise AND logic operations to achieve quasi-stochastic multiplication operations. The resulting bitstream is then compared with the FP8 and the FP64 baseline multiplication results, as shown in Figure~\ref{fig:benchmark}(c,d). As 119 distinctive positive numbers are mapped in FP64 format, the multiplication benchmark spans (119 x 119) multiplication operations to cover all possible combinations. BP10 multiplication results report an average absolute error of 0.30\%, while FP8 demonstrates 0.03\%. The multiplication precision of FP8 is enhanced by the exponential data representation, especially when multiplying smaller values.
However, BP10 shows a significant reduction ($\sim$4.0x) in the average absolute error for the multiplication results (0.30\%) compared to the data mapping (1.19\%), due to positive and negative errors.

The accuracy of $MatMul$ benchmarking is evaluated to explore the effects of accumulating errors through different matrix dimensions. Two square input matrices (NxN) are randomly generated, with the ideal $MatMul$ results generated in FP64 format. The two input matrices are mapped to FP8 and BP10 formats. Then, the aforementioned multiplication operations take place, while a fully binary accumulation approach is performed for BP10, to achieve the highest accuracy~\cite{agwa2023digital}. The binary accumulation results are shifted and scaled into probability values for a fair comparison with FP8 results. This $MatMul$ accuracy analysis is conducted using Frobenius norm calculations. The Frobenius norm of the ideal \added{FP64 }matrix A and the relative Frobenius error are computed as shown in Equation~\ref{equ:1}, where $\hat{A}$ is the test matrix mapped to FP8 and BP10.
\begin{align}\label{equ:1}
\| A \|_F = \sqrt{\sum_{i=1}^{m} \sum_{j=1}^{n} |a_{ij}|^2} && \text{Error} = \frac{\| A - \hat{A} \|_F}{\| A \|_F}
\end{align}




Figure~\ref{fig:benchmark}(e,f) illustrates the relative Frobenius error and its average across the whole $MatMul$ benchmarking space. The benchmarking space in Figure~\ref{fig:benchmark}(e) covers various matrix dimensions from 4x4 to 512x512, thereby studying the impact of matrix scalability on the accuracy metric. For better coverage, the same matrix dimension ($N$x$N$) is repeated 100 times with randomly generated input data. Therefore, each benchmarking dimension ($N$x$N$) covers as many cases as possible. Then, we calculate the average error for each dimension to represent the average relative Frobenius error. 
During the accumulation of individual multiplication results, the positive and negative errors of BP10 tend to offset each other, leading to a reduction in overall error. Consequently, as matrix dimensions increase, the error cancellation becomes more effective, resulting in a significant decrease in BP10 $MatMul$'s average Frobenius error, from 9.42\% for 4x4 matrices to 1.81\% for 512x512 matrices. This improvement in accuracy starts to saturate at larger dimensions, showing minimal improvement with further increases in matrix dimensions. 
We further expand the $MatMul$ accuracy experiment to matrix dimensions that are commonly used in the AI models in Figure~\ref{fig:benchmark}(f), starting from 128x64 matrices multiplied by 64x128 matrices.
The 2.16\% average relative Frobenius error (compared to the ideal FP64) presents BP10 as a promising AI data format. 

Although BP has a 10-bit format to logically present 10 different probabilities from 0.0 to 0.9, only 8 bits effectively contribute to the multiplication operation. As the left-most bit is always zero for all right-biased bitstreams, the bit-wise SC multiplication (AND) with this bit is always zero; therefore, it doesn't contribute to the output result. Similarly, the right-most bit is always zero for all left-biased bitstreams, and the bit-wise SC multiplication (AND) with this bit is always zero. Consequently, removing right-most and left-most bits of the two complementary datasets does not affect the SC multiplication results, considering that output bitstreams are still scaled by 10. This reduces the number of actual bits to 8, improving the code density of BP by 25\%. To achieve higher throughput, area efficiency, and energy efficiency, our proposed architecture, OISMA, has a compressed hardware interpretation of 8-bit for BP10. All possible SC multiplication combinations were verified to be identical for BP8 and BP10. For example, a right-biased bitstream X = 0.3 will be P\textsubscript{0.3}(BP8)~$= 00001110$ instead of P\textsubscript{0.3}(BP10)~$= 0000011100$, while a left-biased bitstream Y = 0.6 will be P\textsubscript{0.6}(BP8)~$= 11111100$ instead of P\textsubscript{0.6}(BP10)~$= 0111111000$. The SC multiplication output is P\textsubscript{0.2}(BP8)~$= 00001100$, which has two ones and is interpreted as 0.2 (not 0.25), similar to the result in Figure~\ref{fig:sc_and}.

\section{OISMA ARCHITECTURE}
\begin{figure*}[!t]
    \centering
    \includegraphics[width=\textwidth]{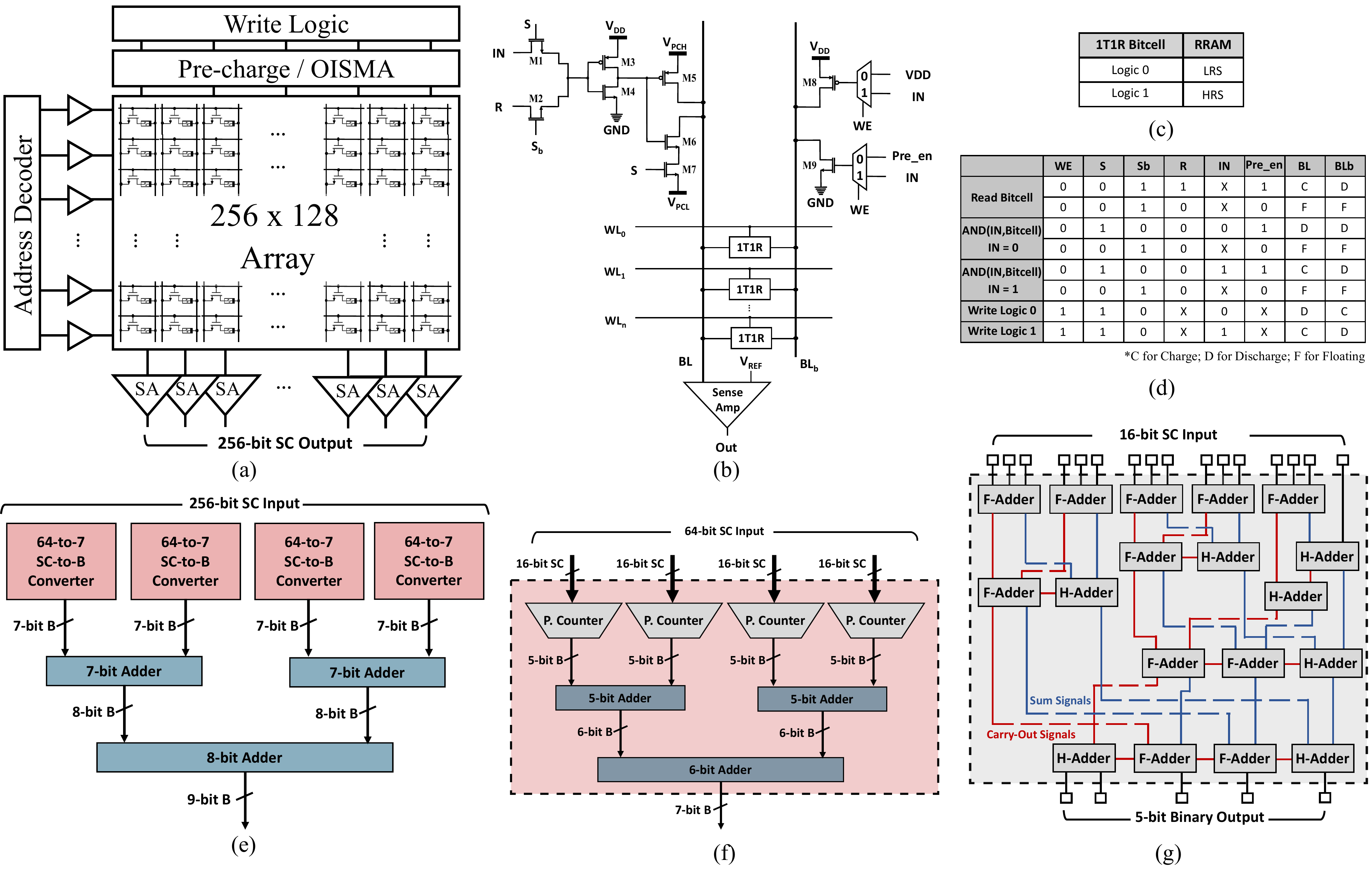}
    \caption{(a) Block diagram of 4KB OISMA array featuring 256 columns and 128 rows of $1T1R$ bitcells. OISMA control logic is integrated within the pre-charging/discharging logic. (b) Column-wise transistor-level implementation 
    (c) RRAM resistance states. (d) Control signals for OISMA column\added{ and the corresponding bitline BL/BLb status. The Read and AND computation operations take two phases (pre-Charge/Discharge and then Floating \& Sensing)}. 
    (e-\deleted{f}\added{g}) Block diagram of: accumulation periphery of an OISMA array with 256-bit SC input and 9-bit binary output; 64-bit SC to 7-bit binary converter; 
    and parallel counter with 16-bit SC input and 5-bit binary output. }
    \label{fig:schematic}
\end{figure*}


OISMA is an On-the-fly In-memory Stochastic Multiplication Architecture that incorporates two main computing units to achieve the Multiply-and-Accumulate (MAC) operation in a hybrid computing domain: 1) a digital 1T1R memory array that is capable of performing a massively parallel in-situ stochastic multiplication operations between inputs and wordlines, and 2) an accumulation periphery of digital Parallel Counters and adder trees for the parallel accumulation of the output stochastic multiplication bitstreams. Figure~\ref{fig:schematic}(a) presents the block diagram of a 4KB OISMA memory array that features 256 columns and 128 rows (256Cx128R) of $1T1R$ bitcells. 
This array utilizes an emerging non-volatile RRAM technology to build a $1T1R$ bitcell structure. It supports three different operations: (1) read bitcells from the wordline, (2) bit-wise AND operation between bitcells and inputs for stochastic multiplication, and (3) write bitcells. The address decoder activates one wordline at a time, while bitlines are controlled on a column basis. The OISMA control logic is integrated within the pre-charging/discharging logic. The write circuitry for RRAM programming re-utilizes part of the OISMA logic to improve the area footprint. Each column exports a single output bit, which is interpreted by a single-ended sense amplifier using a reference voltage. Based on the pre-determined reference voltage at design time, the sense amplifier interprets the discharging rate of the bitline $BL$ into $Logic~1$ or $Logic~0$. 
The detailed transistor-level implementation of an OISMA column is depicted in Figure~\ref{fig:schematic}(b), and the corresponding control signals are highlighted for each operation in Figure~\ref{fig:schematic}(d).


For conventional read operation, the process is divided into two phases (pre-Charge/Discharge and then Floating \& Sensing). First, transistor $M2$ passes the read enable signal $R$, and activates $M5$ to pre-charge $BL$ to $V_{PCH}$. $BL_{b}$ is always pre-discharged to $GND$ through transistor $M9$. In the second phase, both $BL$ and $BL_{b}$ are disconnected and held floating, therefore preparing for the sensing process. During the floating phase, the address decoder activates one wordline $WL$ to observe a charge flow through the activated $1T1R$ bitcell. The digital RRAM has only two states: High Resistance State ($HRS$) for $Logic~1$ and Low Resistance State ($LRS$) for $Logic~0$. The two different resistance states affect the $BL$'s discharging rate, allowing the sense amplifier to distinguish clearly between $HRS$ and $LRS$ to produce an amplified output. When the bitcell has $HRS$, the discharging rate is too slow, and the sense-amplifier reads $Logic~1$. In contrast, if the bitcell has $LRS$, the discharging rate is too fast and the sense-amplifier reads $Logic~0$.

Regarding the multiplication operation \added{in Figure 3(d) }between the input $IN$ and the data stored in the RRAM-based bitcell, the pre-charge level of $BL$ is controlled by the input data instead of the read control signal $R$: if $ IN = 1$ the $BL$ is pre-charged to $V_{PCH}$, while if $ IN = 0$ the $BL$ is pre-discharged to a predetermined low voltage $V_{PCL}$. This is achieved by passing the $IN$ signal through transistor $M1$. Apart from the pre-charging phase, the multiplication operation shares the same floating and sensing process as the read operation. If the input $IN$ is $Logic~1$, the $BL$ is already pre-charged as if it is a normal read operation. The sense amplifier reads the data stored in the bitcell as either $Logic~1$ or $Logic~0$, thereby performing an $AND$ operation. If the input $IN$ is $Logic~0$, the $BL$ is already pre-discharged to a low voltage $V_{PCL}$ and the sense amplifier always reads $Logic~0$, also giving the result of an $AND$ operation.

For the write operations, $BL$ and $BL_{b}$ are set to specific voltages determined by the logic state of input $IN$ and the polarity alignment of the RRAM device. The operation reuses the OISMA logic circuits on $BL$ to obtain a constant voltage level, the same as the first phase of multiplication, where the voltage level is determined by input $IN$. At the $BL_{b}$, an additional transistor $M8$ is applied to allow $BL_{b}$ to be charged to a high voltage level when it is needed, with two multiplexers managing the switch of control signals from read and multiplication modes to write mode.


Having OISMA logic per column enables forwarding many inputs from another memory array in parallel to all bitcells of the wordline, thereby utilizing the full bandwidth. Therefore, massively parallel stochastic multiplication operations are accomplished within a single cycle. The multiplication output bitstreams are accumulated by the accumulation periphery, incorporating parallel counters and adder trees as shown in Figure~\ref{fig:schematic}(e-g). 
Each 256Cx128R OISMA array features four 64-to-7 converters with further binary data accumulation using two 7-bit adders and one 8-bit adder.
A 64-bit SC to 7-bit binary converter is made up of four parallel counters with an adder tree of two 5-bit adders and one 6-bit adder.
The schematic of 16-bit SC to 5-bit binary parallel counter consists mainly of \added{1-bit }full adders (F-Adder) and half-adders (H-Adder), which are carefully connected to achieve the accumulation functionality with minimal hardware overhead. \added{Further accumulation is done through ripple-carry adders for a lower energy footprint.}  
This near-memory accumulation periphery performs two primary tasks: 1) data conversion from the SC domain to the binary domain, and 2) the accumulation required to achieve the matrix-multiplication functionality. Moreover, this accumulation periphery inherently implies spatial data compression; therefore, keeping computations as near as possible to the memory while significantly reducing the output data transmission from OISMA. This multi-functionality of the accumulation periphery, including parallel counters and adder trees, amortizes the cost of data conversion.



Regarding digital IMC, the conventional bitline computing approach must align the input matrix $(Multipliers)$ within the same memory array of the weight matrix $(Multiplicands)$~\cite{Al-Hawaj-iscas2020,Al-Hawaj-hpca2023,agwa_iscas2022, disca}, which implies not only higher input data redundancy but also capacity limitations and more intra-memory data transmission. In contrast, OISMA stores inputs in one array and weights in the other arrays (still within the same memory), avoiding any input data redundancy while fully utilizing the capacity of the memory array. 
Thus, OISMA is equipped with two main principles: 1) maximizing data utilization, and 2) minimizing data movement. These two principles are inherited from Systolic Arrays (SA), where data is held stationary with minimal movement to improve energy efficiency by maximizing data re-usability among the different processing elements~\cite{DiP_2024}. OISMA incorporates 3D stationary approach: 1) Weight Stationary (WS): all weight matrices are stored in the memory without any movement, 2) Input Stationary (IS): input vector $X_i$ is read once and used many times before reading a new input vector $X_{i+1}$, and 3) Partial Sum/Output Stationary (OS): SC bitstreams are accumulated and compressed spatially within the OISMA array.

\begin{figure}[!t]
    \centering
    \includegraphics[width=0.8\linewidth]{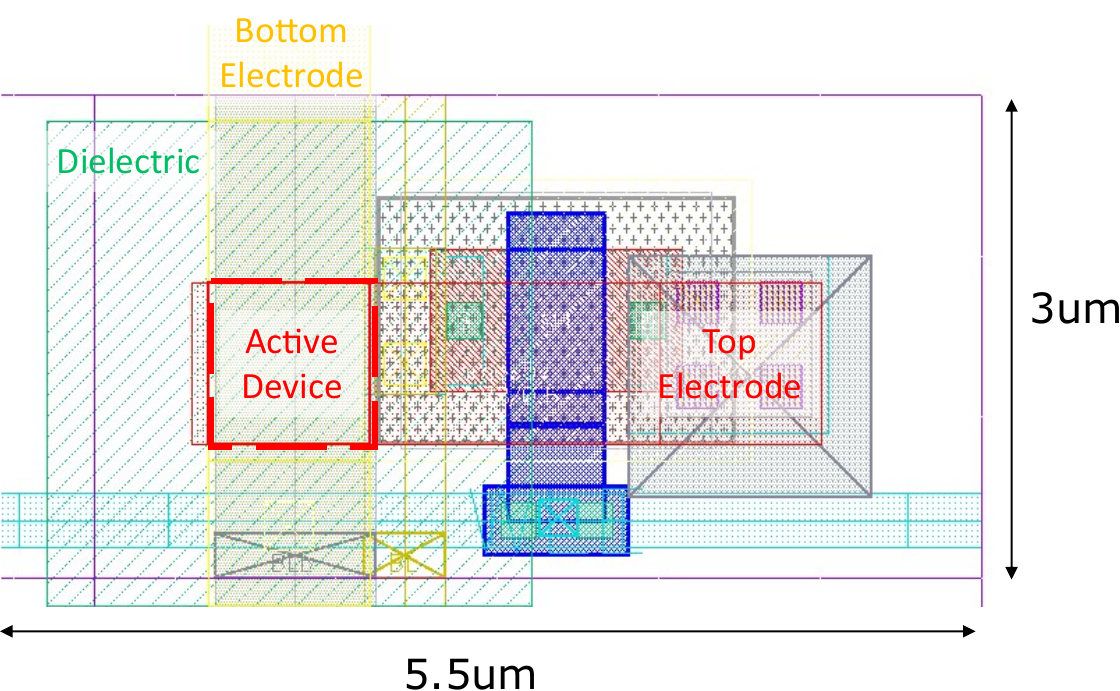}
    \caption{Layout of 1T1R bitcell. Horizontally-placed top electrode connects to transistor through via, forming a metal-insulator-metal stack with the vertically placed bottom electrode.}
    \label{fig:layout}
\end{figure}

The OISMA architecture was prototyped with commercial 180nm technology, where the memory array was implemented through a custom IC design flow, and the accumulation periphery was built using a digital ASIC flow. 
Using the Back-End-of-Line approach, RRAM devices are fabricated in-house with a metal-insulator-metal configuration of Au/TiO2/Pt~\cite{TCASI_design_1}.
Figure~\ref{fig:layout} shows the layout of the 1T1R bitcell with a dimension of 3um x 5.5um. The RRAM top electrode is placed horizontally and intersects with a vertically routed bottom electrode. A dielectric layer is placed between the two electrodes; the overlapped area forms an active RRAM device. The bitcell pitch size is determined by RRAM due to the design constraint set by the in-house RRAM fabrication.
To meet the RRAMs' write voltage requirements, a 5V transistor is used in the bitcell.
However, both read and multiplication operations are conducted under a 1.6V power supply to minimize energy consumption. 

\begin{figure}[!t]
    \centering
    \includegraphics[width=\linewidth]{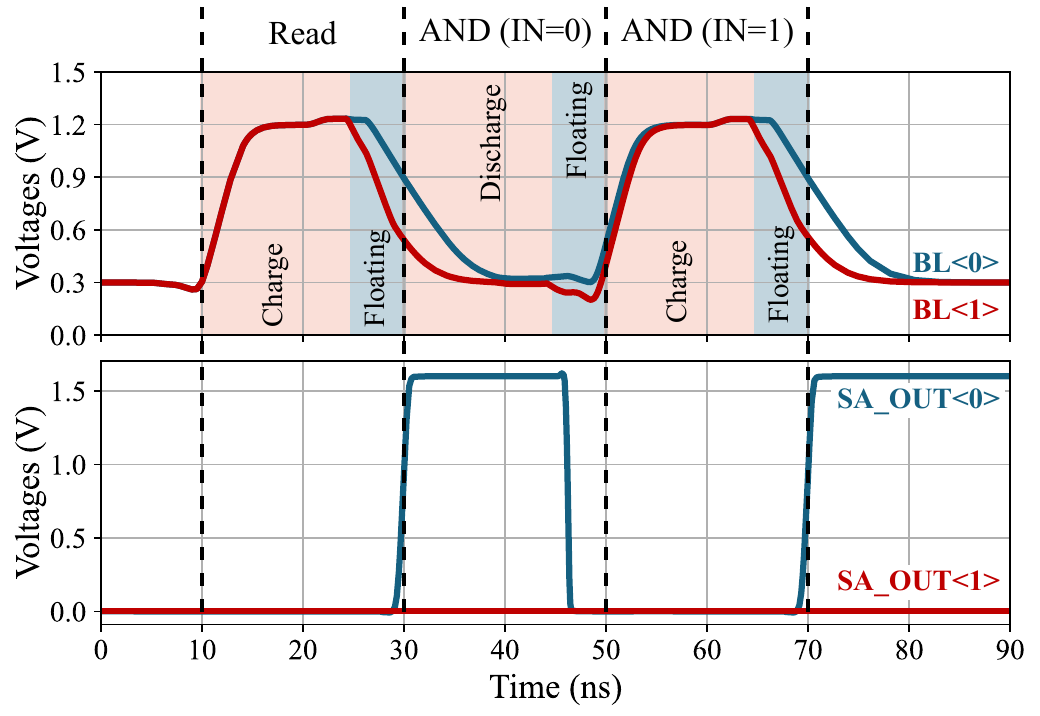}
    \caption{Dynamic simulation of the OISMA array for the read and SC multiplication operations. Two scenarios are displayed where the bitcell stores $Logic~1$ at $BL<0>$ (in blue) and $Logic~0$ at $BL<1>$ (in red). 
    }
    \label{fig:sim}
\end{figure}

The in-house RRAM is simulated by a data-driven Verilog-A model from measured RRAM data~\cite{TCAD_verilogA}. 
The simulation defines RRAM $LRS$ at 112k\textOmega\space as $Logic~0$, and $HRS$ at 8.04M\textOmega\space as $Logic~1$.
RRAM utilizes binary data storage (high and low resistance states), which inherently provides a larger sensing margin against non-idealities such as resistance drift and cycle-to-cycle variation.
Figure~\ref{fig:sim} depicts the dynamic simulation of the OISMA read and SC multiplication (AND) operations. It shows two cases of bitlines associated with two bitcells where $Logic~1$ is for $BL$\textless0\textgreater\space (blue) and $Logic~0$ is for $BL$\textless1\textgreater\space (red).

OISMA operations have a total maximum delay of 20ns, while the sense amplifier generates the output $SA\_OUT$ at the end of the floating \& sensing phase. The pre-charge/discharge phase consumes $\sim$14ns (70\% of the total delay). This is relevant to the low power supply and also the considerable capacitive load, including parasitic effects along the $BL$ spanning 128 bitcells. 
Allowing enough time for $BL$ to charge completely is essential to avoid incorrect readings by the sense amplifier. In addition to ensuring that $BL$ is fully charged, further adjustments were made to the pre-charge/discharge voltage levels, therefore reducing the redundant charge transfer. \added{The sense amplifier compares the voltage at $BL$ against an adjustable input reference voltage $Vref$. }Consequently, the voltage range of $BL$ became 0.3V (representing $Logic~0$) to 1.2V (representing $Logic~1$)\added{ while Monte Carlo simulations show a sensing margin of 184mV  for the worst-case PVT variations}. The compressed voltage swing reduces energy consumption during the multiplication operations, due to the quadratic effect of voltage scaling. This compressed voltage range enables a faster recovery from the pre-discharged state back to the pre-charge state if the input changes from $Logic~0$ to $Logic~1$, leading to faster operations.

\begin{figure}[!t]
    \centering
    \includegraphics[width=\linewidth]{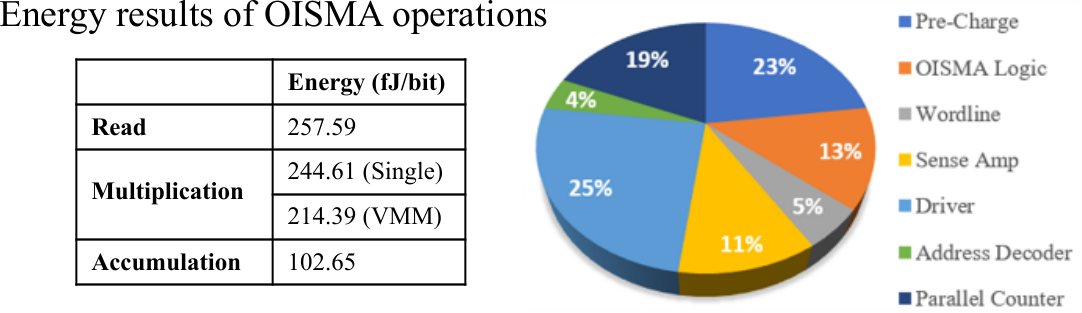}
    \caption{Energy results and breakdown of OISMA operations at 50MHz. Continuative multiplication in VMM consumes less energy per bit as inputs are kept stationary. }
    \label{fig:energy}
\end{figure}

\begin{figure}[!t]
    \centering
    \includegraphics[width=0.8\columnwidth]{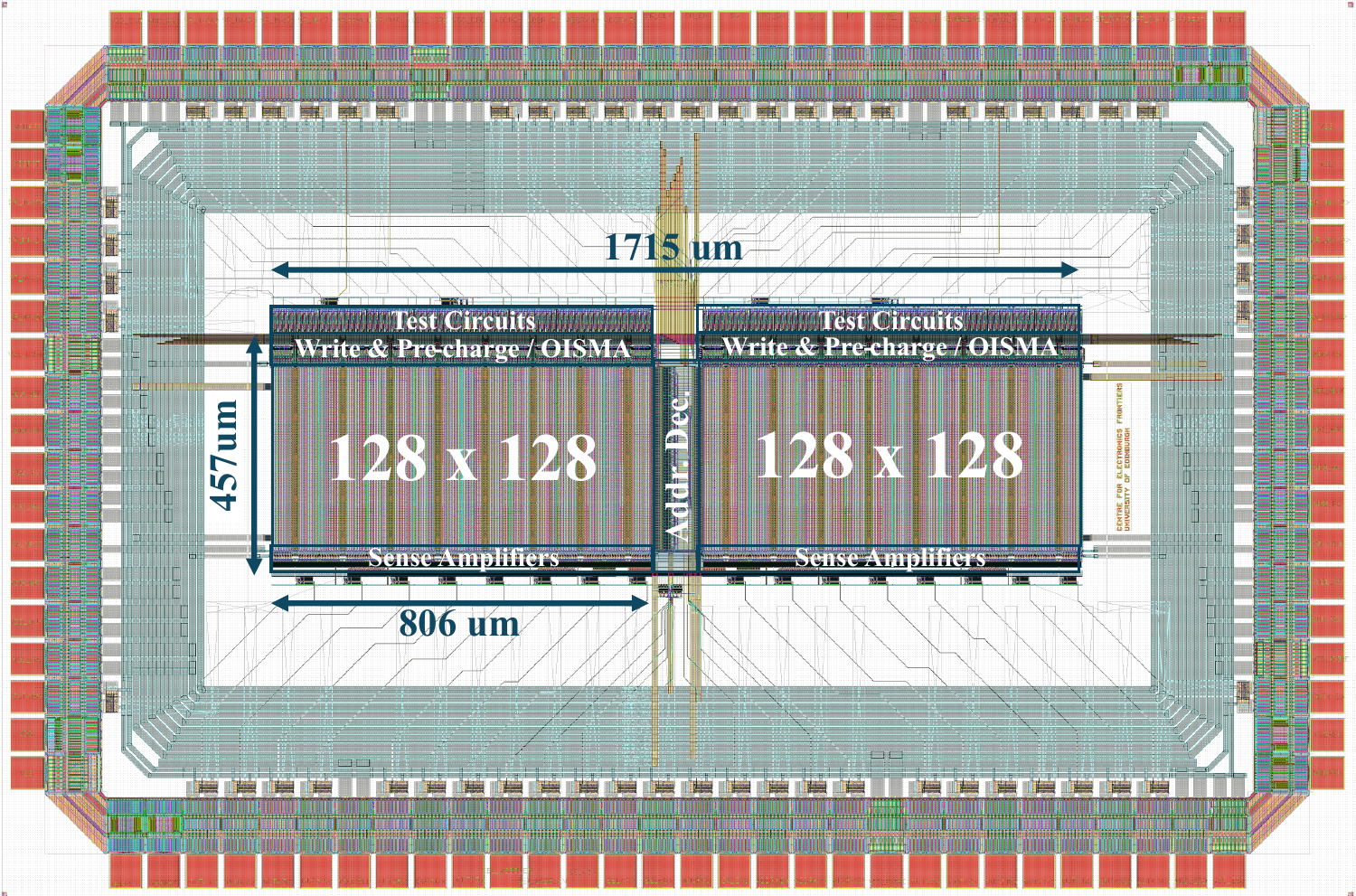}
    \caption{Chip layout of OISMA memory array.}
    \label{fig:chip}
\end{figure}

During the floating phase of the read operation, $BL$\textless1\textgreater\space discharges faster than $BL$\textless0\textgreater\space, indicating that the activated bitcell of $BL$\textless1\textgreater\space has an $LRS$ RRAM device compared to the bitcell of $BL$\textless0\textgreater\space, which has an $HRS$ RRAM device. Consequently, $SA\_OUT$\textless0\textgreater\space shows a high voltage level ($Logic~1$) at $Time$ = 30ns, while $SA\_OUT$\textless1\textgreater\space continues at low voltage level ($Logic~0$) at the same time. If the OISMA multiplication operation has $IN$ at $Logic~0$, the sense amplifier produces a low voltage output representing $Logic~0$, regardless of the data stored by the relevant RRAM device. If $IN$ equals $Logic~1$, the $BL$ discharges only if the stored data is $Logic~0$ represented by an $LRS$ RRAM device, as shown by $SA\_OUT$\textless1\textgreater\space, while $SA\_OUT$\textless0\textgreater\space has $Logic~1$ due to the $HRS$ of the relevant RRAM device. The amplified results of the OISMA multiplication operations for $BL$\textless0\textgreater\space and $BL$\textless1\textgreater\space are observed at $Time$ = 70ns.

The energy results of the 4KB OISMA array are recorded in Figure \ref{fig:energy}, for a 50 MHz operational frequency. The energy measurements are based on randomly generated data sets with balanced logic state distribution of 50\% $Logic~1$ and 50\% $Logic~0$. 
The energy figures reflect the average consumption of eight iterations for each operation.
It is observed that OISMA multiplication operations consume less energy than read operations due to: 1) the compressed voltage range, and 2) higher probability of lower switching activities for the output results, as SC multiplications usually generate more $Logic~0$ outputs than read operations. 

\newcommand{\greyrule}{\arrayrulecolor{black!30}\midrule\arrayrulecolor{black}}

\begin{table*}[!t]
\caption{Comparison of key parameters and performance metrics \deleted{between}\added{for} OISMA \added{simulated at 180nm and projected at 22nm} \deleted{and}\added{versus} other state-of-the-art systems.}
\label{tab:compare}
\centering
\resizebox{\textwidth}{!}{%
\begin{tabular}{lllllll}
\hline
 &
  ISCAS'20 \cite{Al-Hawaj-iscas2020} &
  TC'23 \cite{eidetic2023} &
  ISSCC'25 \cite{Zhiheng_isscc2025} &
  ISSCC'24 \cite{Taihao_isscc2024} &
  ISSCC'25 \cite{deqi_isscc2025} &
  OISMA \\ \hline
Memory Technology &
  SRAM, 28nm &
  SRAM, 22nm &
  SRAM, 28nm &
  RRAM, 22nm &
  STT-MRAM, 22nm &
  RRAM, 180nm \\
Data Format &
  INT8 / INT32 &
  INT8 / FP16 &
  INT8 / FP8 / FP16 &
  BF16 / FP16 &
  INT8 &
  BP8 \\
Voltage &
  0.9V &
  - &
  0.62V - 0.9V &
  0.7V - 0.8V &
  0.65V - 0.8V &
  1.2V, 1.6V, 2V\textsuperscript{b}\\
Frequency &
  900MHz &
  1.5GHz &
  100MHz - 525MHz &
  167MHz - 208MHz &
  38MHz - 56MHz &
  50MHz (372MHz\textsuperscript{a}) \\
Power &
   - &
  235.7W &
   - &
   - &
  1.22mW &
  4.06mW (0.30mW\textsuperscript{a}) \\
   \midrule
\begin{tabular}[c]{@{}l@{}} Throughput \\ (TOPS) \end{tabular} &
  \begin{tabular}[c]{@{}l@{}}0.076 (INT8)\\ 0.006 (INT32)\end{tabular} &
  \begin{tabular}[c]{@{}l@{}}175.5 (INT8)\\ 41.8 (FP16)\end{tabular} &
   - &
  \begin{tabular}[c]{@{}l@{}}0.86 (BF16)\textsuperscript{d}\\ 0.78 (FP16)\textsuperscript{d}\end{tabular} &
  0.128\textsuperscript{e} &
  \begin{tabular}[c]{@{}l@{}}0.003\\ (0.024\textsuperscript{a})\end{tabular} \\ \hline
 \begin{tabular}[c]{@{}l@{}} Energy Efficiency\\ (TOPS/W) \end{tabular} &
  \begin{tabular}[c]{@{}l@{}}0.116 (INT8)\\ 0.009 (INT32)\end{tabular} &
  \begin{tabular}[c]{@{}l@{}}0.745 (INT8)\\ 0.177 (FP16)\end{tabular} &
  \begin{tabular}[c]{@{}l@{}}115.0$\sim$43.2 (INT8)\textsuperscript{c}\\99.7$\sim$37.4 (FP8)\textsuperscript{c}\\ 51.6$\sim$15.1 (FP16)\textsuperscript{c}\end{tabular} &
  \begin{tabular}[c]{@{}l@{}}31.2 (BF16)\textsuperscript{d}\\ 28.7 (FP16)\textsuperscript{d}\end{tabular} &
  104.5\textsuperscript{e} &
  \begin{tabular}[c]{@{}l@{}}0.789\\ (79.22\textsuperscript{a})\end{tabular} \\ 
OISMA Improvement &
 \begin{tabular}[c]{@{}l@{}}  683x \\  8,802x \end{tabular} &
 \begin{tabular}[c]{@{}l@{}} 106x \\ 448x \end{tabular} &
  \begin{tabular}[c]{@{}l@{}}
  0.69x$\sim$1.83x \\
  0.79x$\sim$2.12x \\
  1.54x$\sim$5.25x \\
  \end{tabular} &
 \begin{tabular}[c]{@{}l@{}} 2.54x \\ 2.76x \end{tabular} &
  0.76x &
  - \\ \hline
\begin{tabular}[c]{@{}l@{}}Area Efficiency\\ (TOPS/mm$^2$) \end{tabular} &
  \begin{tabular}[c]{@{}l@{}}0.069 (INT8)\\ 0.006 (INT32)\end{tabular} &
  \begin{tabular}[c]{@{}l@{}}0.659 (INT8)\\ 0.157 (FP16)\end{tabular} &
  \begin{tabular}[c]{@{}l@{}}0.72$\sim$3.81 (INT8)\textsuperscript{c}\\ 0.62$\sim$3.30 (FP8)\textsuperscript{c}\\ 0.46$\sim$2.44 (FP16)\textsuperscript{c}\end{tabular} &
  \begin{tabular}[c]{@{}l@{}}0.104 (BF16)\textsuperscript{d}\\ 0.095 (FP16)\textsuperscript{d}\end{tabular} &
  0.036\textsuperscript{e} &
  \begin{tabular}[c]{@{}l@{}}0.004 \\ (3.28\textsuperscript{a})\end{tabular} \\ 
OISMA Improvement &
  \begin{tabular}[c]{@{}l@{}}48x \\ 547x \end{tabular} &
  \begin{tabular}[c]{@{}l@{}}5x \\ 21x \end{tabular} &
  \begin{tabular}[c]{@{}l@{}}
  4.56x$\sim$0.86x \\
  5.29x$\sim$0.99x \\
  7.13x$\sim$1.34x \\
  \end{tabular} &
  \begin{tabular}[c]{@{}l@{}} 32x \\  35x \end{tabular} &
  91x &
  - \\  \hline
\end{tabular}%
}
\parbox{0.98\textwidth}{%
    \footnotesize
    \vspace{0.5mm}
    \textsuperscript{a}Normalized to 22nm using DeepScaleTool\cite{stillmaker2017scaling}\cite{sarangi2021deepscale}
    \textsuperscript{b}Bitline at 1.2V, wordline and sense amplifier at 2V, other parts at 1.6V.
    \textsuperscript{c}Explored static/dynamic sparsity (up to 85\%), measured at 0.62V, 100MHz.
    \textsuperscript{d}Input and weight pre-alignd mantissa, 50\% input sparsity.
    \textsuperscript{e}Measured at 0.8V, 50\% input sparsity.
}

\end{table*}

For the Vector-Matrix Multiplication (VMM) benchmarking, the input vectors are kept stationary and multiplied consequently with the different weight rows stored in OISMA's wordlines. 
This input stationary technique maximizes the data utilization and minimizes the switching activity. Therefore, the VMM benchmarking mode shows lower energy consumption by 12.4\% than the single benchmarking mode, where input vectors are changed every cycle. On the other hand, the accumulation periphery, including parallel counters and adder trees, was implemented using standard cells at the same technology node (180nm) with a 1.6V supply voltage. The energy results show an average energy consumption of 102.65 fJ/bit. Consequently, the average energy consumption of OISMA MAC operation is 347.26 fJ/bit, and using a compressed 8-bit BP format results in 2.778 pJ/MAC at 180nm technology. 
While the area of the accumulation peripherals is 20485.606 µm$^2$, the OISMA array chip depicted in Figure~\ref{fig:chip} occupies a core area of 1715 µm x 457 µm for two sub-arrays, excluding the test circuits. This means that the total effective computing area is 0.804241 mm$^2$. The design has two 128 x 128 sub-arrays that share the same address decoder positioned at the center. Each 128 x 128 sub-array has a dimension of 806 µm x 457 µm. 

\section{PERFORMANCE ANALYSIS}


Table~\ref{tab:compare} compares OISMA with state-of-the-art IMC architectures using different performance metrics, including energy efficiency, area efficiency (silicon computational density), and peak throughput (which is subject to the architecture's scale). 
At 180nm technology, OISMA consumes an average power of 4.06 mW at 50 MHz with a 4 KB memory capacity, thanks to the voltage and frequency scaling techniques applied to reduce power consumption, in addition to data-stationary techniques that reduce the switching activities. The 4 KB OISMA array has a peak throughput of 3.2 GOPS, while the goal is to scale up OISMA to a 1 MB engine, leading to a peak throughput of 819.2 GOPS at 50 MHz. Consequently, OISMA, equipped with a compressed 8-bit version of BP, has an energy efficiency of 0.789 TOPS/W and an area efficiency of 3.98 GOPS/mm$^2$, which represents the computational density of OISMA, attributed to the negligible area overhead represented by OISMA logic.

For a fair comparison, we normalize the results of OISMA and other state-of-the-art IMC architectures to 22nm technology using DeepScaleTool with scaling factors presented by \cite{stillmaker2017scaling}\cite{sarangi2021deepscale}.
Table~\ref{tab:compare} covers various state-of-the-art designs, including digital binary-based IMC using SRAM for only integers \cite{Al-Hawaj-iscas2020}, or integers and floating points \cite{eidetic2023, Zhiheng_isscc2025}, in addition to IMC architectures using emerging technologies such as RRAMs for floating point \cite{Taihao_isscc2024}, and STT-MRAM\cite{deqi_isscc2025} for integer operations. 
At a scaled 22nm technology, OISMA outperforms the state-of-the-art architectures with energy efficiency of 79.22 TOPS/W and area efficiency of 3.28 TOPS/mm$^2$, considering a dense $MatMul$ scenario. 
While dense OISMA achieves comparable energy efficiency as the 50\% input sparsity architecture in \cite{deqi_isscc2025}, it achieves 91x higher computational density. 
Dense OISMA still performs comparably to extreme sparsity IMC scenarios shown in \cite{Zhiheng_isscc2025}, improving energy efficiency by up to 5.25x, and area efficiency by up to 7.13x. 
Compared to dense IMC architectures, OISMA outperforms \cite{eidetic2023} (modeling results) and \cite{Al-Hawaj-iscas2020} (post-layout results) by at least 106x and 683x, respectively, for energy efficiency. Regarding area efficiency (computational density), OISMA outperforms \cite{eidetic2023} and \cite{Al-Hawaj-iscas2020} by at least 5x and 48x, respectively.
\begin{figure}[!b]
    \centering
    \includegraphics[width=0.9\linewidth]{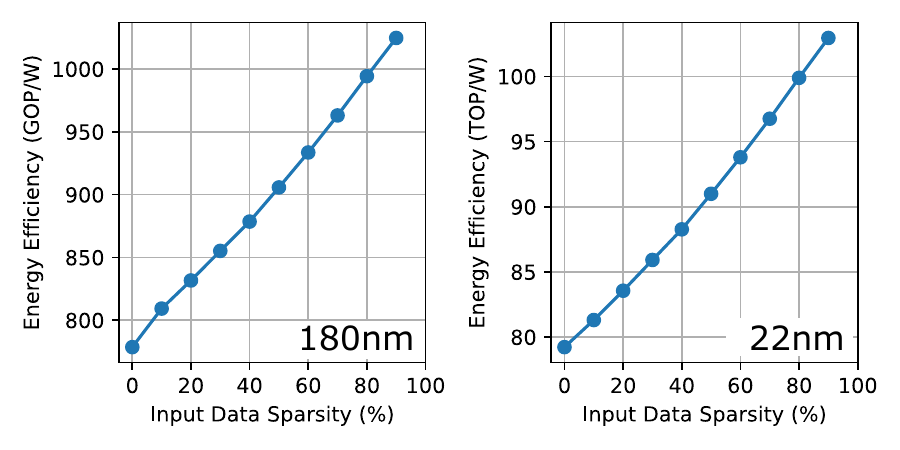}
    \caption{Energy efficiency for BP8 OISMA MAC operation against various input sparsity measured at 180nm and estimated at 22nm.}
    \label{fig:sparsity}
\end{figure}

While OISMA targets dense $MatMul$ scenarios, Figure \ref{fig:sparsity} shows the effect of data sparsity on improving the energy efficiency of OISMA.
Under 50\% input sparsity, the measured energy efficiency is 905.67GOPS/W, representing a 15\% improvement over the dense input cases.

Therefore, OISMA promotes state-of-the-art digital IMC capabilities by an average of two orders of magnitude improvement in energy efficiency and one order of magnitude improvement in area efficiency, without inheriting any design complexity, circuitry uncertainty, or cross-domain bottlenecks of the analog IMC architectures.

\section{CONCLUSION}

This paper presents an in-memory stochastic computing architecture, OISMA, which converts memory read operations into in-situ stochastic multiplication operations with a negligible cost. An accumulation peripheral logic accumulates the multiplication bitstreams to accomplish the matrix multiplication functionality. OISMA, equipped with a compressed Bent-Pyramid 8-bit format, achieves a relative Frobenius error of 1.81\% (for 512x512 matrix dimensions), compared to the 64-bit double precision floating point format. Therefore, OISMA offers a comparable accuracy to the floating-point computing for matrix multiplication, while significantly reducing its computational complexity. At 180nm technology, a 4 KB 1T1R OISMA architecture achieves 0.789 TOPS/W and 3.98 GOPS/mm$^2$ for energy and area efficiency, respectively. OISMA inherits the same scalability benefits of digital systems; therefore, scaling OISMA to 22nm using DeepScaleTool \cite{stillmaker2017scaling,sarangi2021deepscale} shows a significant improvement of two orders of magnitude in energy efficiency and one order of magnitude in area efficiency, compared to its digital counterparts. 
Our future work is to scale up OISMA to a 1MB OISMA engine, while scaling down the technology to 22nm. Quasi-stochastic formats with shorter bitstreams will also be investigated. 


\end{document}